\documentclass[%
  a4paper,
  amsfonts,amssymb,amsmath,
  reprint,
  superscriptaddress,        % <-- add this
  showkeys,nofootinbib,twoside
]{revtex4-2}
\usepackage[english]{babel}
\usepackage[utf8]{inputenc}
\usepackage{amsthm}
\usepackage{siunitx}
\usepackage[version=4]{mhchem}
\usepackage{comment}
\usepackage{graphicx}
\usepackage{subcaption}
\usepackage[colorinlistoftodos, color=green!40, prependcaption]{todonotes}
\usepackage{lineno}
\usepackage{amsthm}
\usepackage{mathtools}
\usepackage{physics}
\usepackage{xcolor}
\usepackage{graphicx}
\usepackage[left=23mm,right=13mm,top=35mm,columnsep=15pt]{geometry} 
\usepackage{adjustbox}
\usepackage{placeins}
\usepackage[T1]{fontenc}
\usepackage{lipsum}
\usepackage{csquotes}
\usepackage{siunitx}
\usepackage[font=small, justification=justified, labelfont=bf]{caption}

\usepackage[pdftex, pdftitle={Article}, pdfauthor={Author}]{hyperref} % For hyperlinks in the PDF
\usepackage{ragged2e}
\begin{document}
\title{Bright Telecom Spin-Photon Interface in Silicon Photonics}

%\begin{comment}
\author{Carolina Crosta}
\thanks{These authors contributed equally to this work.}
%\email{c.crosta3@campus.unimib.it}
\affiliation{Dipartimento di Scienza dei Materiali, Universit\`a di Milano-Bicocca and BiQute, via R. Cozzi, 20125 Milano, Italy}

\author{Amirehsan Alizadehherfati}
\thanks{These authors contributed equally to this work.}
%\email{herfati@umd.edu}
\affiliation{Institute for Research in Electronics and Applied Physics and Joint Quantum Institute, University of Maryland, College Park, Maryland 20742, USA}
\affiliation{Department of Electrical and Computer Engineering, University of Maryland, College Park, MD 20742, USA}

\author{Purbita Purkayastha}
%\email{herfati@umd.edu}
\affiliation{Institute for Research in Electronics and Applied Physics and Joint Quantum Institute, University of Maryland, College Park, Maryland 20742, USA}
\affiliation{Department of Physics, University of Maryland, College Park, Maryland 20742, USA}

\author{Kyu-Young Kim}
\affiliation{Institute for Research in Electronics and Applied Physics and Joint Quantum Institute, University of Maryland, College Park, Maryland 20742, USA}
\affiliation{Department of Electrical and Computer Engineering, University of Maryland, College Park, MD 20742, USA}
\affiliation{Department of Physics, Ulsan National Institute of Science and Technology, Ulsan 44919, Republic of Korea}

\author{Jasvith Raj Basani}
%\email{jasvith@umd.edu}
\affiliation{Institute for Research in Electronics and Applied Physics and Joint Quantum Institute, University of Maryland, College Park, Maryland 20742, USA}
\affiliation{Department of Electrical and Computer Engineering, University of Maryland, College Park, MD 20742, USA}

\author{Chang-Min Lee}
\affiliation{Institute for Research in Electronics and Applied Physics and Joint Quantum Institute, University of Maryland, College Park, Maryland 20742, USA}
\affiliation{Department of Electrical and Computer Engineering, University of Maryland, College Park, MD 20742, USA}

\author{Fabio Pezzoli}
\email{fabio.pezzoli@unimib.it}
\affiliation{Dipartimento di Scienza dei Materiali, Universit\`a di Milano-Bicocca, BiQute and INFN-LNL, Via R. Cozzi, 55, Milano, 20125, Italy}

\author{Edo Waks}
\email{edowaks@umd.edu}
\affiliation{Institute for Research in Electronics and Applied Physics and Joint Quantum Institute, University of Maryland, College Park, Maryland 20742, USA}
\affiliation{Department of Electrical and Computer Engineering, University of Maryland, College Park, MD 20742, USA}

%\end{comment}
%\date{\today} % Leave empty to omit a date

\begin{abstract}
Silicon is an attractive host for scalable quantum photonics, but the absence of bright telecom-band emitters with optically addressable spin states has limited its use for spin-photon interfaces. Here we demonstrate the Al1-center, an aluminum--carbon defect in silicon, as a bright waveguide-integrated single-photon emitter with a ground-state spin. Using isotopically purified silicon-on-insulator nanophotonic devices, we isolate individual Al1-centers and observe high-purity single-photon emission with $g^{(2)}(0)=0.04$ without background subtraction. Time-resolved photoluminescence spectroscopy reveals a fast excited-state lifetime of \SI{135}{ns}, nearly an order of magnitude shorter than the benchmark provided by the well-studied T-center. Resonant photoluminescence excitation measurements further resolve the zero-phonon transition and reveal a narrow homogeneous linewidth reaching 47 MHz, threefold narrower than the T-center under comparable temperature. Through magneto-optical spectroscopy, we resolve the spin-dependent transitions of the bound-exciton manifold and achieve spin-selective optical pumping, fulfilling the prerequisite for quantum state initialization and readout. These results establish the Al1-center as a bright telecom-band spin-photon interface in silicon photonics and introduce a promising platform for integrated quantum networks.
\end{abstract}

\maketitle

\section{Introduction}

Silicon is a promising host for quantum photonics because it enables large-scale integration of a wide range of optical components, including waveguides \cite{zhang2022ultralow,vlasov2004losses}, modulators \cite{han2023slow,liu2004highspeed}, detectors \cite{lischke2021ultra,ahn2007highperformance}, and nanophotonic resonators \cite{bogaerts2012silicon,novick2023highbandwidth}. Silicon photonics is also compatible with commercial semiconductor foundries, allowing the fabrication of large-scale photonic circuits with high-density of components in a compact footprint \cite{thomson2016roadmap,qiang2018largescale}. These advantages have motivated extensive efforts to develop silicon-based quantum photonic technologies \cite{labonte2024integrated,wang2020integrated}. Despite this progress, silicon quantum photonics remains constrained by the lack of efficient on-chip deterministic quantum light sources and optically active spin qubits.

Silicon color centers have emerged as a promising solution to this bottleneck. Several color centers, including the G-center \cite{redjem2020single}, W-center \cite{baron2022detection}, $\mathrm{C_i}$ \cite{jhuria2024programmable} and T-center \cite{higginbottom2022optical}, have been isolated at the single-defect level and shown to emit single photons. Among these centers, only the T-center supports a spin ground state that can serve as a quantum memory, which is essential for quantum networking \cite{azuma2023quantum} and distributed quantum computing \cite{main2025distributed}. The T-center has demonstrated promising electron and nuclear spin coherence times \cite{bergeron2020silicon}, as well as remote spin entanglement \cite{afzal2024distributed} and quantum gates with nearby $^{29}\mathrm{Si}$ nuclear spin qubits \cite{song2026entanglement}. However, the T-center suffers from a fundamental limitation on excited state lifetime of approximately \SI{1}{\micro\second}, intrinsically capping the emitter's photon emission rate. Thus, silicon still lacks an optically active spin qubit that is both bright and capable of supporting long coherence times through a stable spin ground state.

Recently, a high-throughput computational search identified the Al1 point defect as a viable solution \cite{xiong2024computationally}. This emitter, comprising an Al--C substitutional defect, was predicted to support the largest transition dipole moment among T-center-like defects. The Al1-center therefore represents a potential spin qubit that could combine high brightness with long spin coherence times. However, this prediction is awaiting experimental demonstration, and detailed investigations are needed to enrich the present understanding of this defect. To date, experimental investigations of the Al1-center have been restricted to ensemble photoluminescence \cite{noonan1974aluminum, irion1988aluminum, irion1989photoluminescence}. Unlocking their potential as a spin-photon interface requires isolating individual defects to directly validate their recombination properties and confirm the presence of a spin ground state

Here, we demonstrate an efficient spin-photon interface in silicon operating in the telecom S-band. We successfully isolate single Al1-centers and report a 20-fold enhancement in brightness compared to the T-center. Second-order correlation measurements demonstrate that the Al1 defect emits single photons with exceptional purity, exhibiting second order correlation of $g^{(2)}(0) = 0.04$ without background subtraction due to its high single-photon emission rate. Time-resolved lifetime measurements show an excited state lifetime of \SI{135}{\nano\second}, nearly an order of magnitude shorter than that of the T-center. Two-color photoluminescence excitation measurements reveal a narrow homogeneous linewidth of 47 MHz, representing a threefold reduction in the optical dephasing compared to T-centers under similar device and temperature conditions. Finally, by resolving the spin-dependent transitions under an external magnetic field, we demonstrate spin-selective optical pumping, a critical prerequisite for initializing and reading out a quantum memory. These results establish the Al1 defect as a bright, optically active spin qubit with potential applications in quantum networking and distributed quantum computing.

\section{Physical System and Device Overview}

\begin{figure}
    \centering
    \includegraphics[width=\columnwidth]{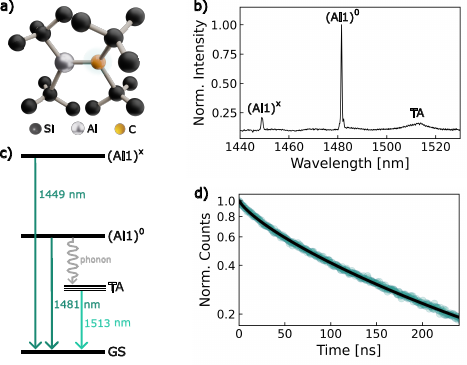}
     \caption{ \textbf{Atomic structure and ensemble optical characterization.} \textbf{a)} Atomic structure of the Al1-center in silicon ((Al--C)\textsubscript{Si}); \textbf{b)} Photoluminescence emission of an Al1-center ensemble showing a sharp peak at 1481 nm [(Al1)\textsuperscript{0}], corresponding to the zero-phonon line, a broad band feature centered at 1513 nm (TA) associated to the transverse acoustic phonon sideband, and a second peak at 1449 nm [(Al1)\textsuperscript{x}] that we ascribe to recombination from the second excited state of the defect; \textbf{c)} Summary of proposed electronic structure of the Al1-center; \textbf{d)} Time-resolved photoluminescence measurement of the ensemble (dots) and corresponding stretched exponential fit (solid line), which provides $\tau = 128.5 \pm 1.2$ ns and $\beta = 0.752 \pm 0.006$.}
    \label{fig1}
\end{figure}

\textbf{Figure \ref{fig1}a} shows the expected atomic structure of the Al1-center \cite{xiong2024computationally}. The structure is assigned to a substitutional acceptor–carbon complex, (Al--C)\textsubscript{Si}, in which an aluminum atom and a carbon atom share a silicon lattice site. This defect is T-center-like because it is structurally and electronically analogous to the T center (C--C--H)$_\mathrm{Si}$. In both cases, the relevant in-gap defect state is primarily localized on the unsaturated carbon atom. Because its p orbital contains a single unpaired electron, the neutral defect has a doublet, spin-1/2, ground state. The predicted point-group symmetry of the Al1-center is C$_\mathrm{2v}$, which is consistent with the rhombic-I symmetry inferred from stress spectroscopy \cite{irion1989photoluminescence}. This makes the Al1-center more symmetric compared to the T-center, where the presence of the hydrogen atom lowers the symmetry to C$_\mathrm{1h}$ \cite{safonov1996interstitial}. As a consequence, while T-center ensembles can distribute among twelve magnetically non-equivalent orientations \cite{clear2024optical}, ensembles of Al1-centers are expected to occupy six magnetically non-equivalent orientations. This reduction in the orientation complexity simplifies the engineering of magnetic fields required for addressing multiple spins across wafer-scale devices.

We synthesize Al1-centers using controlled sequential carbon and aluminum implantation into a purified silicon-on-insulator (SOI) platform, followed by thermal annealing. The \hyperlink{methods}{Methods} section provides additional details on the generation conditions of the defect. 

To efficiently collect emission from single Al1 defects, we employ a tapered nanobeam design (see \textbf{Figure \ref{fig:SEM_beams}}) \cite{lee2023high}. The \hyperlink{methods}{Methods} section provides additional details on the design and fabrication of the nanobeam structure. For all the results shown in the following, we collect the emission from the beam using a lensed fiber, while exciting the top surface of the sample. From reflectivity measurements we determine the nanobeam - fiber coupling efficiency to reach 70\% for the devices utilized in this experiment. More detail about the experimental setup is provided in \hyperlink{methods}{Methods}.

\section{Photoluminescence Characterization}
\label{sec:PL}

We initially characterize the sample at \SI{8}{\kelvin} using photoluminescence spectroscopy under above-band excitation at \SI{800}{\nano\meter}. \textbf{Figure \ref{fig1}b} shows the measured ensemble photoluminescence spectrum, whose main feature is a narrow emission line at 1481.4 nm, labeled as (Al1)\textsuperscript{0}. This emission line is consistent with the zero-phonon line of the Al1-center previously reported \cite{noonan1974aluminum, irion1988aluminum}. We do not observe this peak in samples that have not been implanted with Al, validating that this element is essential to generate this emission. In addition, the spectrum exhibits a weaker transverse acoustic phonon replica at 1513 nm, labeled as TA, which is in agreement with the phonon replica wavelength identified by \cite{irion1988aluminum, irion1989photoluminescence}. We further confirm this assignment via photoluminescence excitation spectroscopy by selectively exciting the zero-phonon line and collecting the phonon replica emission. A third peak is visible at a shorter wavelength of 1449 nm, labeled as (Al1)\textsuperscript{x}. The appearance of this line is consistently correlated with the presence of the zero-phonon emission across different beams, and its temperature behavior closely resembles the (Al1)\textsuperscript{0} trend (see \textbf{Figure \ref{fig:PL(T)_ensemble}}). The (Al1)\textsuperscript{x}--(Al1)\textsuperscript{0} energy distance of $\sim$\SI{18}{\milli\electronvolt} well matches the $15 \pm 5$ \SI{}{\milli\electronvolt} extracted from Al1-center ensembles in bulk silicon \cite{irion1989photoluminescence} as the first dissociation energy of the defect, which is attributed to the population of an excited luminescent state. Thus, we ascribe the (Al1)\textsuperscript{x} line to recombination from the Al1-center second excited state. The suggested level structure of the point defect is illustrated in \textbf{Figure \ref{fig1}c}. Unlike the T-center, whose narrow 1.76 meV TX\textsubscript{0}–TX\textsubscript{1} splitting severely restricts its operational temperature range \cite{bergeron2020silicon}, the Al1-center boasts an excited-state separation nearly an order of magnitude larger, providing robust immunity against thermal depopulation.

Driven by the need for higher photon emission rates than existing silicon spin systems, we perform time-resolved photoluminescence measurements to evaluate the excited-state dynamics. \textbf{Figure \ref{fig1}d} displays the decay of emission as a function of time measured on the ensemble (dots). The solid line is a fit to the experimental data using a stretched-exponential model, yielding a characteristic excited-state decay time of $\tau = 128.5 \pm 1.2~\mathrm{ns}$ and a stretching parameter of $\beta = 0.752 \pm 0.006$. This decay time is nearly an order of magnitude shorter than the reported T-center lifetime \cite{bergeron2020silicon}, consistent with the predicted stronger electric dipole moment of the Al1-center \cite{xiong2024computationally}. The non-unity value of $\beta$ suggests a distribution of local atomic and strain environments among the centers contributing to the measured ensemble emission.
 
\begin{figure}
    \centering
    \includegraphics[width=\columnwidth]{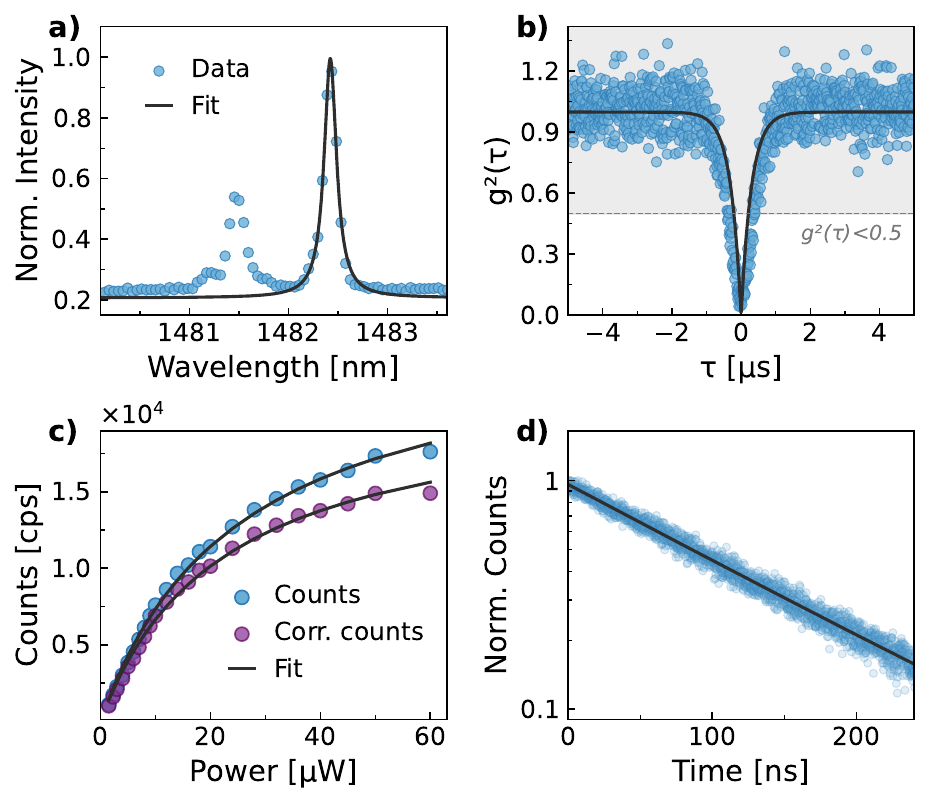}
    \caption{\textbf{Optical characterization of a single Al1-center.} \textbf{a)} Photoluminescence spectrum (dots) and corresponding Lorentzian fit (solid line) of the zero-phonon line at 1482.437 nm. The extracted linewidth is 20.2 ± 0.7 GHz; \textbf{b)} Second-order autocorrelation measurement under continuous wave 800 nm pumping with power of 3 $\mu$W (dots) and corresponding fit with \textbf{Equation \ref{eq:g2}} (solid line). The extracted zero-delay value is $g^{(2)}(0) = 0.04 \pm 0.02$; \textbf{c)} Count rate as a function of CW 800 nm incident power without and with $g^{(2)}(0)$ correction (blue and purple dots, respectively) and two-level system fit with \textbf{Equation \ref{eq:saturation}} (solid line). Fitting of the corrected counts provides $P_{\mathrm{sat}} = 25 \pm 1 \; \mu$W and $I_{\mathrm{sat}} = (2.14 \pm 0.05) \times 10^4$ cps; \textbf{d)} Time-resolved photoluminescence measurement under pulsed 780 nm excitation with repetition rate of 2.5 MHz (dots) and corresponding single exponential fit (solid line), extracting $\tau = 135 \pm 2$ ns.}
    \label{fig2}
\end{figure}

Having established the ensemble optical signature of the Al1-center, we perform spatially resolved photoluminescence measurements across the nanobeam array and spectrally screened each device to identify isolated zero-phonon lines. Owing to the stochastic nature of defect formation under the present implantation and annealing conditions, the local Al1-center density varies between nanobeams, producing ensemble emission in some devices and discrete lines from individual centers in others. \textbf{Figure~\ref{fig2}a} shows the photoluminescence spectrum of one isolated emitter, with a zero-phonon line centered at 1482.44~nm. To validate that this emission originates from a single-photon emitter, we perform second-order correlation measurements. \textbf{Figure \ref{fig2}b} shows the measured second-order correlation $g^{(2)}(\tau)$ collected from the zero-phonon line (see \hyperlink{methods}{Methods} for experimental details). The correlation exhibits near-complete suppression of coincidences at zero time delay, indicating emission from a single quantum emitter with high single-photon purity. To quantify the purity of the single-photon emitter, we fit the experimental data with an analytical antibunching model, shown as the solid line (see Appendix \ref{PL_APP}). From this fit, we extract a zero-delay value $g^{(2)}(0)=0.04 \pm 0.02$. Obtaining such exceptionally high single-photon purity without the need for cavity Purcell enhancement demonstrates the viability of the isolated Al1 defect as a highly efficient, standalone quantum emitter.

Beyond single-photon purity, a practical spin-photon interface must deliver high photon flux; therefore, we directly benchmark the saturation brightness of the isolated Al1-center against existing metrics. \textbf{Figure \ref{fig2}c} shows the measured intensity as a function of the above-band pump power. We plot both the raw measured counts (blue dots), as well as the $g^{(2)}$-corrected count rate (purple dots), which extracts the true single photon emission rate by removing background that artificially increases the brightness but does not correspond to emission from the Al1-center. The corresponding power-dependent $g^{(2)}(0)$ can be found in \textbf{Figure \ref{fig:g2(P)}}, and exhibits high single-photon purity even at emitter saturation power. To quantify the saturated emitter brightness, we fit the saturation curve to a two-level system model (see Appendix \ref{PL_APP}). From this fit, we extract a saturation pump power of $25 \pm 1\; \mu$W and a saturated count rate after background and $g^{(2)}(0)$ correction (see \textbf{Equation \ref{saturation_curve_correction}}) of $(2.14 \pm 0.05) \times 10^4$ cps. We compare this saturated count rate with emission from T-centers measured in the same nanobeam geometry \cite{lee2023high}. This comparison shows a 20-fold increase in brightness for the Al1-center, indicating that the Al1-center is substantially brighter than the T-center under comparable device conditions.

To confirm that this enhanced brightness stems from fundamental transition dynamics, we measure the excited-state lifetime of the single Al1 defect. \textbf{Figure~\ref{fig2}d} shows the emission intensity decay as a function of time. The decay fits very well to a single exponential decay model (black solid line), with an excited-state lifetime of $\tau = 135 \pm 2~\mathrm{ns}$. Contrarily to the ensemble measurement, which requires a stretched-exponential model, the single-center decay is well described by a single exponential, further supporting the assignment of this isolated spectral line to an individual Al1-center. By exhibiting nearly an order of magnitude faster transition than the T-center, the Al1 defect leverages its stronger transition dipole moment to deliver a fundamentally brighter spin-photon interface.

\section{Photoluminescence Excitation Measurements}
\label{sec:PLE}

Having established bright, pure single-photon emission and a fast decay from an isolated Al1-center, we next probe the resonant optical response of this defect. For this purpose, we use a second identical sample cooled to a base temperature of \SI{4.2}{\kelvin} in a magneto-optical cryostat capable of applying magnetic fields of up to \SI{9}{\tesla}. We perform photoluminescence excitation and spin measurements on a single Al1-center within this sample. \textbf{Figure \ref{fig:p2nb4_single}} provides additional characterization measurements on this center.

\begin{figure*}
    \centering
    \includegraphics[width=\textwidth]{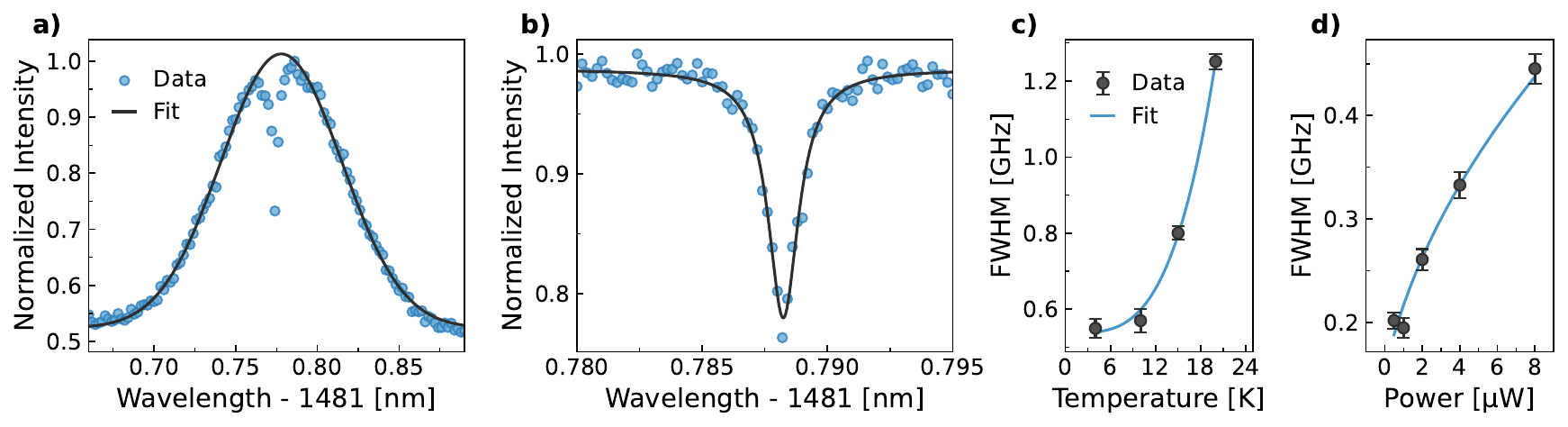}
    \caption{\textbf{Transient spectral hole burning measurements on a single Al1-center.} \textbf{a)} Two-color photoluminescence excitation spectroscopy of the zero-phonon line with a resolution of 2 pm (dots) and Gaussian fit (solid line), extracting FWHM = $10.85 \pm 0.08$ GHz. The incident power of the scanning laser is 0.5 $\mu$W, while the pumping laser power is 1.5 $\mu$W; \textbf{b)} Magnification of the hole burning dip obtained with 0.2 pm resolution and both scanning and pumping laser power of 0.5 $\mu$W (dots), and corresponding Lorentz fit (solid line) computing FWHM = $0.202 \pm 0.008$ GHz; \textbf{c)} FWHM behavior as a function of sample temperature with both scanning and pumping laser power of 5 $\mu$W (dots) and corresponding fit (solid line) according to \textbf{Equation \ref{eq:holeburning_T}}; \textbf{d)} FWHM behavior as a function of pumping laser power (dots) and fit using \textbf{Equation \ref{eq:holeburning_P}} (solid line). Scanning power is kept constant at 0.5 $\mu$W and temperature is fixed at \SI{4.2}{K}.}
    \label{fig3}
\end{figure*}

We conduct resonant excitation on the (Al1)\textsuperscript{0} transition by tuning a narrow-linewidth laser near 1481.8 nm, and collecting the corresponding phonon replica emission at 1513 nm through a bandpass filter. The resulting intensity as a function of the scanning laser wavelength is shown in \textbf{Figure \ref{fig:Single_PLE}}, which provides an inhomogeneous linewidth of $10.85 \pm 0.08$ GHz by fitting the spectrum with a Gaussian profile. This value is significantly broader than the lifetime-limited linewidth, indicating broadening mechanisms such as local strain, electric-field disorder, or spectral diffusion in the sample \cite{zhang2025laser}. We attribute the presence of such mechanisms to a residual implantation-induced damage of the crystal lattice due to incomplete recovering of the crystalline order during the annealing step of the center generation process. 
This interpretation is consistent with the broad defect-related photoluminescence background observed in the same nanobeam, which is displayed in \textbf{Figure~\ref{fig:PL_lattice_disorder}}. Further optimization of the implantation and annealing conditions will be required to improve the spectral stability of the transition (see \textbf{Appendix \ref{PL_APP}} for further discussion).

To bypass local disorder and uncover the true coherence properties that dictate the emitter's suitability for quantum networks, we extract the homogeneous linewidth via two-tone photoluminescence excitation spectroscopy. In this measurement, the pump laser is fixed at the center of the inhomogeneous peak, while a second, weaker probe laser is scanned across the same spectral region. The resulting spectrum as a function of scanning laser wavelength is shown in \textbf{Figure \ref{fig3}a}, where the solid line is the best Gaussian fit to the single tone inhomogeneous profile, presented as comparison. When the probe laser is resonant with the spectral class addressed by the pump, the probe-induced emission is suppressed, producing a spectral hole. This suppression arises from saturation of the emitter by the pump laser. Once the emitter drifts away, the emission recovers to the inhomogeneous spectrum trend. \textbf{Figure \ref{fig3}b} shows a higher-resolution measurement of the spectral hole, and its corresponding Lorentzian best fit to the dip (solid line). For this specific experimental condition we extract a full width at half maximum of $0.202 \pm 0.008$ GHz from the fit. 

To investigate the mechanism governing homogeneous broadening, we repeat the measurement over a span of sample temperatures. \textbf{Figure \ref{fig3}c} shows the extracted full width at half maximum of the spectral hole as a function of temperature. Specifically, the dip width increases monotonically with temperature. We fit this trend using a power-law broadening model (\textbf{Equation \ref{eq:holeburning_T}}) from which we extract a scaling factor of $\sim T^{3.5 \pm 0.4}$, where $T$ is the sample temperature. This behavior is close to the $T^3$ scaling predicted for defect-mediated optical dephasing, in which acoustic phonons modulate long-range interactions between an optical center and nearby defects \cite{hizhnyakov1999optical}. Given the pronounced spectral diffusion of the emitter, this mechanism is probable, although the limited temperature range does not uniquely distinguish it from crossover or mixed dephasing processes.

Finally, to determine the intrinsic homogeneous linewidth and photon coherence, we measure the spectral-hole dip width as a function of pump-laser power while keeping the probe-laser power fixed. The resulting power broadening trend is shown in \textbf{Figure \ref{fig3}d}. By fitting the data with a power-broadening model (\textbf{Equation \ref{eq:holeburning_P}}) and extrapolating to zero pump power, we extract a homogeneous linewidth of $\Delta \omega_{\mathrm{hom}}/2\pi = 47 \pm 8$ MHz for this transition. Despite the local lattice disorder, this value is approximately threefold narrower than the benchmark reported for T-centers at comparable temperatures \cite{zhang2025laser}. This advantage becomes even more stark when benchmarked against theoretical lifetime limits. While our Al1-center broadens from a 1.2 MHz limit to 47 MHz (a factor of $\sim 40$), the T-center degrades from 160 kHz to over 140 MHz, a thousand-fold broadening driven entirely by its severe vulnerability to thermal depopulation of TX\textsubscript{0} \cite{zhang2025laser}.

\section{Spin-Selective Optical Pumping}

\begin{figure*}
    \centering
    \includegraphics[width=\textwidth]{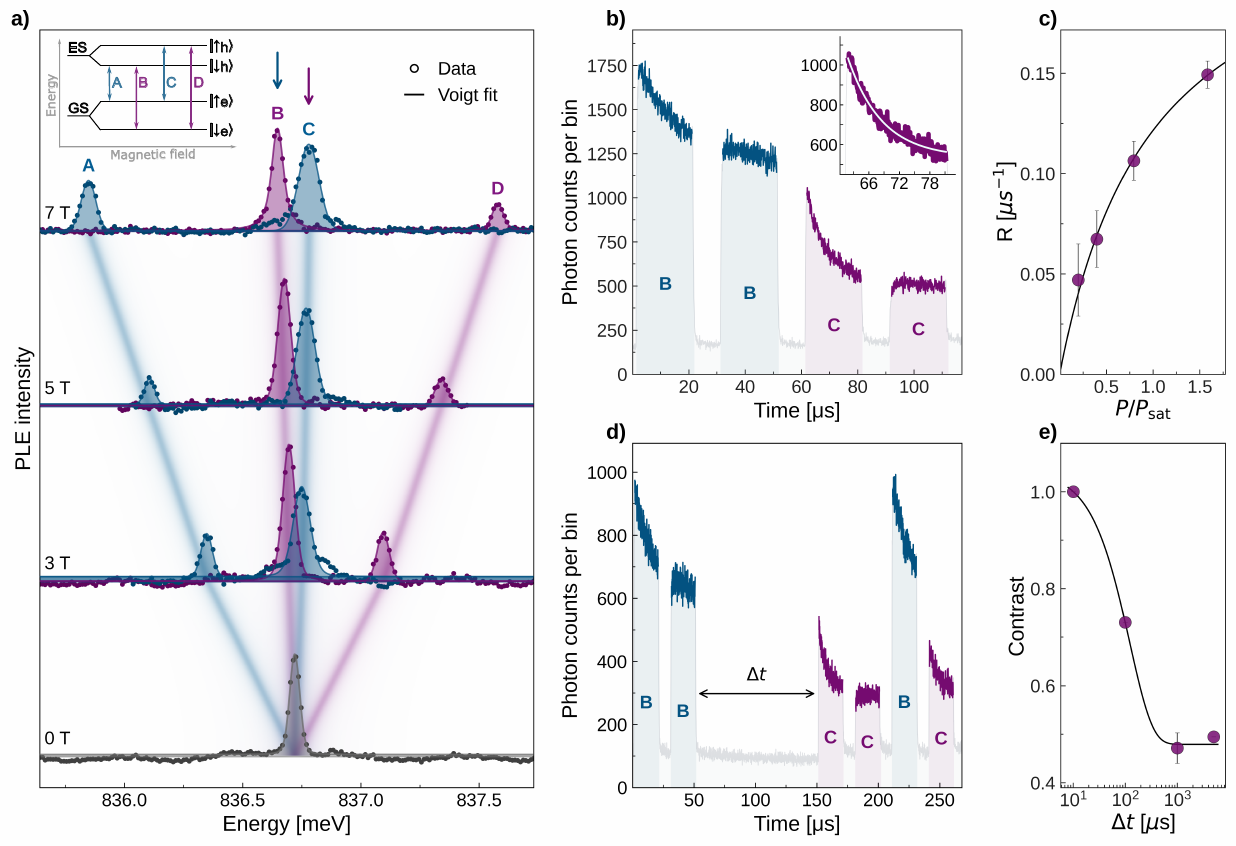}
    \caption{\textbf{Magnetospectroscopy and time-resolved optical pumping of a single Al1-center.} \textbf{a)} Zeeman splitting via two-color photoluminescence excitation spectroscopy. The blue (purple) spectrum is obtained by fixing the pump laser at peak B (C) energy. The inset figure illustrates the level structure of bound excitonic transitions in presence of an external magnetic field, showing four spin-allowed optical transitions; \textbf{b)} Pulse sequence used to demonstrate optical pumping. The first two pulses pump the population into the opposite spin state, while the second pair transfers the population back to the initial state. The difference between the measured B- and C-pulse amplitudes arises from the additional background underneath transition C. Inset: exponential fit of the third pulse decay; \textbf{c)} Power dependence of the optical-pumping rate, proving faster population transfer at higher pump power; \textbf{d)} Pulse sequence used to measure the ground-state spin relaxation time. The first two pulses initialize the population into the opposite spin state. After a variable delay $\mathrm{\Delta t}$, the first probe pulse measures the recovered population, followed by an immediate reference pulse pair; \textbf{e)} Fitting of the extracted contrast with an exponential model to compute the spin-polarization lifetime.}
    \label{fig4}
\end{figure*}

To establish the Al1-center as a viable quantum memory, we probe its spin properties using two-tone resonant measurement in the presence of an external magnetic field up to \SI{9}{\tesla} and perpendicular to the sample surface. \textbf{Figure~\ref{fig4}a} shows the resulting emission from this single Al1-center, where the pump laser is fixed on the main transitions while the probe laser is scanned across the spectral range, exciting selective spin-resolved optical transitions at different magnetic field levels. The magnetic field lifts the degeneracy of the electron and hole spin states, resolving the 4 allowed optical transitions of the bound-exciton manifold, labeled as A--D and shown as an inset. When the pump laser is fixed on transition B (C), emission is recovered only when the probe laser is resonant with transitions A and C (B and D). This behavior is consistent with spin-selective optical pumping into a dark spin state within the four-level system. From the magnetic-field dependence of the four transition energies, we extract the electron and hole $g$-factors. Linear fits to the transition Zeeman splittings yield $g_e = 1.98 \pm 0.02$ and $g_h = 2.30 \pm 0.02$ (see \textbf{Figure \ref{fig:g-factor}}). \textbf{Figure \ref{fig:4centers_PL(B)}} provides additional statistics on effective g-factors observed on other single emitters among the sample. 

We next demonstrate optical initialization and readout of the spin state by means of time-resolved spin selective pumping experiment. To minimize the spectral overlap between B and C transitions, magnetic field is fixed to \SI{7}{\tesla}. We apply a sequence of four \SI{20}{\micro\second} resonant optical pulses, where the first two pulses are resonant with transition B and the second two pulses are resonant with transition C, as shown in \textbf{Figure~\ref{fig4}b}. For each pair, the first pulse exhibits a transient decay in the emission, corresponding to optical pumping out of the addressed spin state, while the second pulse measures the residual background after the population has been transferred to the opposite spin state. 

To quantify spin-selective population transfer, we repeat the measurement while varying the optical power on transition C and keeping the power on transition B fixed. This ensures that the same initial spin polarization is prepared for each measurement, while only the strength of the C optical cycle is changed. For each power, the transient fluorescence decay is fitted with an exponential model (\textbf{Equation \ref{eq:pumping_transient}}), as illustrated by the white solid line in the inset of \textbf{Figure~\ref{fig4}b}, and the extracted optical-pumping rate is plotted in \textbf{Figure~\ref{fig4}c}. The rate initially increases with excitation power and then approaches a constant value, showing that the population-transfer dynamics are limited by the resonant excitation rate at low power and by saturation of the optical transition at high power. Fitting this dependence with the saturation model described by \textbf{Equation \ref{eq:pumping_saturation}} yields a maximum optical-pumping rate of
$\gamma_{\mathrm{p}}^{\mathrm{sat}} = 0.24 \pm 0.05~\mathrm{\mu s}^{-1}$.
Together with the independently measured excited-state lifetime, this rate corresponds to an optical cyclicity of $\eta = 13.7 \pm 2.9$. While integration with optical cavities is ultimately necessary for single-shot readout, the achieved cyclicity is sufficient for high-fidelity all-optical spin initialization without the need for microwave control.

While rapid readout and initialization are vital, the ultimate utility of the Al1-center as a resilient quantum memory hinges on the relaxation of its optically prepared spin state. To measure the longitudinal relaxation time, we use the delayed pump-probe sequence shown in \textbf{Figure~\ref{fig4}d}. Two pulses resonant with transition B first transfer the population into the ground-state level addressed by transition C. After a variable dark interval $\mathrm{\Delta t}$, the first C pulse probes the spin polarization that remains. The fluorescence transient is proportional to the difference between the spin polarization present at the beginning of the pulse and the steady-state polarization produced by C excitation. The following C pulse establishes the background after optical pumping, while the final B--C pair re-prepares and reads out the same polarized state to provide an in-sequence reference. The normalized transient contrast therefore directly measures the decay of the optically prepared spin polarization during the dark interval.

The decay of the optically prepared spin polarization is shown in \textbf{Figure~\ref{fig4}e}. For each dark interval $\mathrm{\Delta t}$, we determine the normalized polarization contrast from the difference between the first C readout pulse and the final reference pulse after background subtraction. As $\mathrm{\Delta t}$ increases, the contrast decreases toward its long-delay value as the ground-state population relaxes toward equilibrium. We fit this dependence using an exponential decay model (see \textbf{Equation \ref{eq:contrast}}), which yields a longitudinal spin-relaxation time of $T_{1}=123\pm27~\mathrm{\mu s}$ at \SI{7}{\tesla} and \SI{4.2}{K}. This value is considerably shorter than the $T_{1}=80\pm30~\mathrm{ms}$ measured for waveguide-integrated T centers near \SI{0.2}{\tesla} at \SI{1}{K} and the lower bound of $T_{1}>16~\mathrm{s}$ reported for bulk T-centers near \SI{0.08}{\tesla} at \SI{1.4}{K} \cite{bergeron2020silicon,deabreu2023waveguide}. The difference is consistent with the much higher magnetic field used in this work. For localized electron spins in silicon, the direct one-phonon relaxation rate is expected to scale approximately as $T_{1}^{-1}\propto B^{5}\coth\!\left(\frac{g_e\mu_{\mathrm{B}}B}{2k_{\mathrm{B}}T}\right)$, because both the spin--phonon coupling and the acoustic-phonon density of states increase with the Zeeman splitting \cite{abragam2012electron,hasegawa1960spin}. The measured value therefore likely reflects enhanced spin--phonon relaxation in the high-field regime rather than an intrinsically short ground-state lifetime. Further investigations are necessary to identify the dominant relaxation mechanism and rule out other possibilities such as laser leakage and interaction with residual defects due to lattice damage.

\section{Conclusion}
\label{sec:develop}

In conclusion, we establish the Al1-center as a bright telecom-band spin–photon interface in silicon. Individual centers exhibit high-purity single-photon emission with $g^{(2)}(0) = 0.04$ and an excited-state lifetime of \SI{135}{ns}, nearly an order of magnitude shorter than that of the T-center and consistent with the predicted stronger optical transition of the Al--C defect. Resonant spectroscopy reveals a homogeneous linewidth of 47 MHz, while magnetospectroscopy and time-resolved measurements resolve the spin-dependent optical transitions and demonstrate spin-selective initialization of the ground-state population.

Further progress will require improved material processing to suppress implantation-induced spectral diffusion and stabilize the optical transition. Narrower inhomogeneous linewidths would enable spin-selective operation at lower magnetic fields, where spin relaxation is reduced, and coherent microwave control becomes more accessible. Combined with cavity enhancement and local registers formed from the intrinsic aluminum and carbon nuclear spins, these advances could support coherent spin control, efficient spin–photon entanglement, and long-lived quantum memories. The integration of bright telecom emission, an optically addressable ground-state spin, and silicon nanophotonics positions the Al1-center as a promising building block for scalable quantum-network nodes.

\section*{Acknowledgments}
The authors acknowledge Yuxi Jiang for valuable discussions. C.C. acknowledges financial support from the Ermenegildo Zegna Founder's Scholarship. The Waks group would like to acknowledge financial support from the National Science Foundation (grant \#ECCS2423788), the Department of Energy (grant \#DESC0026071), and the Air Force Office of Scientific Research (grant \#FA95502310667 and \#FA95502410266).

\section*{DATA AND MATERIALS AVAILABILITY}
All of the data that support the findings of this study are reported in the main text and Appendix. Source data are available from the corresponding authors on reasonable request.

\section*{Competing interests}
The authors declare that they have no competing interests.

\section*{Author contributions} 
C.C., A.A., K.K., F.P., and E.W. conceived the experiment. P.P. fabricated the device. A.A. and C.C. performed the experiment. A.A. and C.C. analyzed the experimental data. C.C., A.A., F.P. and E.W. prepared the manuscript. J.R.B. assisted with preparation of the figures. All authors discussed the results and confirmed the manuscript. E.W. and C.L. supervised the experiment.

\hypertarget{methods}{}
\section*{Methods} %\label{sec:Methods}

\setcounter{figure}{0}
\renewcommand{\thefigure}{M\arabic{figure}}
\setcounter{equation}{0}
\renewcommand{\theequation}{S\arabic{equation}}
\setcounter{subsection}{0}

\subsection*{Al1-center generation} \label{subsec:generation}

\ce{(Al-C)_{Si}} complexes have been generated in purified \ce{^{28}Si}-on-insulator via subsequent implantation at room temperature of C (\SI{38}{\kilo\electronvolt} energy, \SI{7.2e12}{ions/cm^{2}} dose) and Al (\SI{40}{\kilo\electronvolt} energy, \SI{1e12}{ions/cm^{2}} dose). The ion energy has been chosen to produce an ion projected range at a depth of \SI{110}{\nano\meter} from simulations made by The Stopping and Range of Ions in Matter (SRIM) software. Implantation was performed with a \SI{7}{\degree} tilting of the ion beam with respect to the sample surface to reduce channeling phenomena during the implantation process, which would lead to a distorted ion projected range and thus a distorted defect depth and concentration profiles. We employed a \SI{300}{\celsius} post-implantation annealing for \SI{1}{h} to heal the implantation lattice damage and to favor carbon mobility to form the \ce{(Al-C)_{Si}} complex of interest.

\subsection*{Device description} \label{subsec:device}
\begin{figure}
    \centering
    \includegraphics[width=\columnwidth]{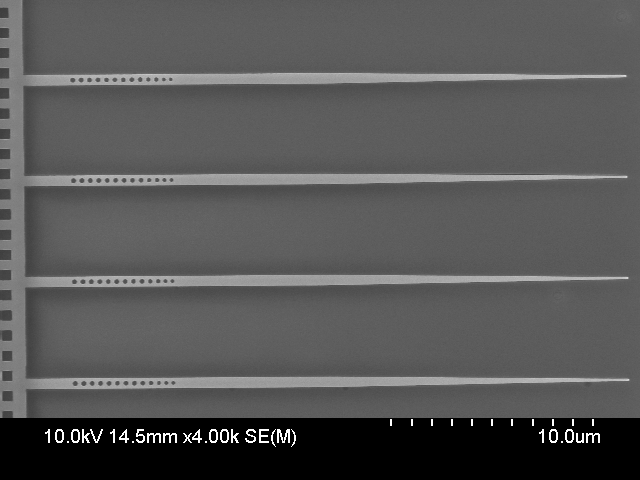}
    \caption{Scanning electron microscope image of the nanobeams.}
    \label{fig:SEM_beams}
\end{figure}

To efficiently collect emission from individual Al1-centers, we use a tapered silicon nanobeam waveguide designed for edge coupling to a lensed fiber, shown in \textbf{Figure \ref{fig:SEM_beams}}. The device consists of a nanobeam of width $b$ connected to an adiabatic taper of length $L_{\mathrm{taper}}$, which reduces the waveguide width to $b_{\mathrm{taper}}$ at the output facet. This taper gradually expands the guided optical mode, improving mode matching to the transverse profile of the lensed fiber used for collection. At the opposite end of the nanobeam, a one-dimensional photonic crystal mirror formed by a periodic array of air holes with radius $r$ and lattice constant $a$ reflects emission back toward the tapered output, thereby directing the collected light preferentially toward the fiber-coupled end.

The taper geometry is designed to couple efficiently to a lensed fiber with numerical aperture $\mathrm{NA}=0.4$. Finite-difference time-domain simulations yield optimized device parameters of $a=\SI{420}{nm}$, $r=\SI{126}{nm}$, $b=\SI{520}{nm}$, $b_{\mathrm{taper}}=\SI{110}{nm}$, and $L_{\mathrm{taper}}=\SI{15}{\micro m}$.

The nanobeam devices are fabricated using electron-beam lithography followed by etching of the silicon-on-insulator substrate. To enable efficient edge collection, selected nanobeams are subsequently transferred to the edge of a separate silicon carrier chip using transfer-print lithography. In this process, a polydimethylsiloxane (PDMS) stamp is used to pick up an individual patterned nanobeam from the source substrate and place it near the edge of the carrier wafer. The resulting suspended geometry allows emission from Al1-centers coupled to the nanobeam mode to be collected directly from the chip edge using a lensed fiber.

\subsection*{Experimental Setup} \label{subsec:setup}
For Photoluminescence characterization, the sample is mounted in a closed-loop cryostat (S50 Montana Instruments) reaching base temperature of \SI{8}{\kelvin}. The sample is excited from top via a 100x objective with NA of 0.70 (Mitutoyo, Plan Apo NIR). Above-band excitation at 780-800 nm is provided by a continuous wave laser (Msquared, Solstis) and a pulsed laser (Picoquant, LDH-D-C-780) with 2.5 MHz repetition rate.

For photoluminescence excitation spectroscopy and optical spin manipulation measurements, the sample is mounted in a closed-loop cryostat (Attocube, Attodry1000) with a base temperature of \SI{4.2}{\kelvin} equipped with a superconducting magnet providing fields up to \SI{9}{\tesla}. A free-space home-built confocal microscope is used for excitation, with a \(0.68\)-NA objective and a focal length of 3 mm. Resonant excitation at the zero-phonon line is provided by narrow band resonant lasers (Santec, TSL 570 and TOPTICA, CTL 1500) and a wavemeter (Bristol 671) to monitor and lock both lasers to the same reference. The resonant lasers pass individually through two polarization-maintaining fiber-coupled acousto-optic modulators (AOM, Brimrose Corp), which are controlled by an arbitrary
waveform generator (Keysight, 33512B).

In all measurements signal is collected through a lensed fiber (working distance of \SI{14}{\micro\meter}) coupled to the silicon nanobeam by measuring the reflected light power, reaching coupling efficiencies of $\sim$70\%. The collected signal from the zero-phonon line is selectively filtered using a tunable fiber-coupled filter (WLPhotonics) with a bandwidth of 0.3 nm. The collected signal from phonon sideband is filtered using a home-built free-space filtering setup containing a bandpass filter (Thorlabs, FBH051520-12) with a bandwidth of 12 nm. For correlation measurements, the collected signal is split by a fiber-coupled \(50{:}50\) beam splitter (Thorlabs). Spectra and detections are acquired with a Princeton Instruments spectrometer comprising a monochromator (600\,g/mm and 150\,g/mm grating) and a 1024 pixel InGaAs camera and two fiber-coupled superconducting single-photon detectors (QuantumOpus) with efficiency of 80\% using a high speed time tagger (PicoQuant, HydraHarp 400 and Multiharp 160).

\appendix
\renewcommand{\sectionautorefname}{Appendix}
\section*{Appendix}
\setcounter{subsection}{0}

\section{Photoluminescence characterization}
\label{PL_APP}
\setcounter{figure}{0}
\renewcommand{\thefigure}{A\arabic{figure}}
\setcounter{equation}{0}
\renewcommand{\theequation}{A\arabic{equation}}

\subsection*{Temperature dependence of the Al1-center emission lines}
\begin{figure}
    \centering
    \includegraphics[width=\columnwidth]{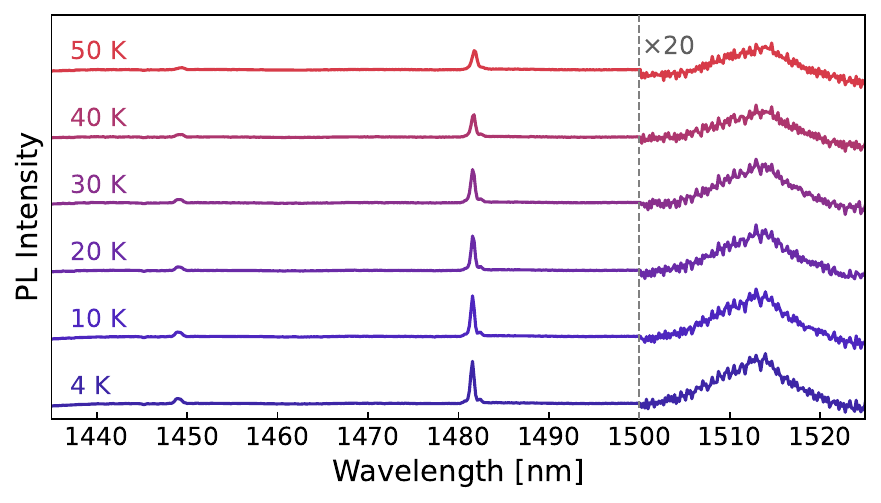}
    \caption{Ensemble photoluminescence spectra as a function of sample temperature.}
    \label{fig:PL(T)_ensemble}
\end{figure}

\textbf{Figure \ref{fig:PL(T)_ensemble}} shows the temperature dependence of an ensemble of Al1-centers in a silicon nanobeam. We observe emission from all three lines associated to the Al1 defect and described in Section \ref{sec:PL}, namely (Al1)\textsuperscript{x} at 1449 nm, (Al1)\textsuperscript{0} at 1481 nm, and the phonon replica at 1513 nm. They all show the same intensity decreasing behavior as the sample temperature is increased.

\subsection*{Implantation induced lattice disorder}

\begin{figure}
    \centering
    \includegraphics[width=\columnwidth]{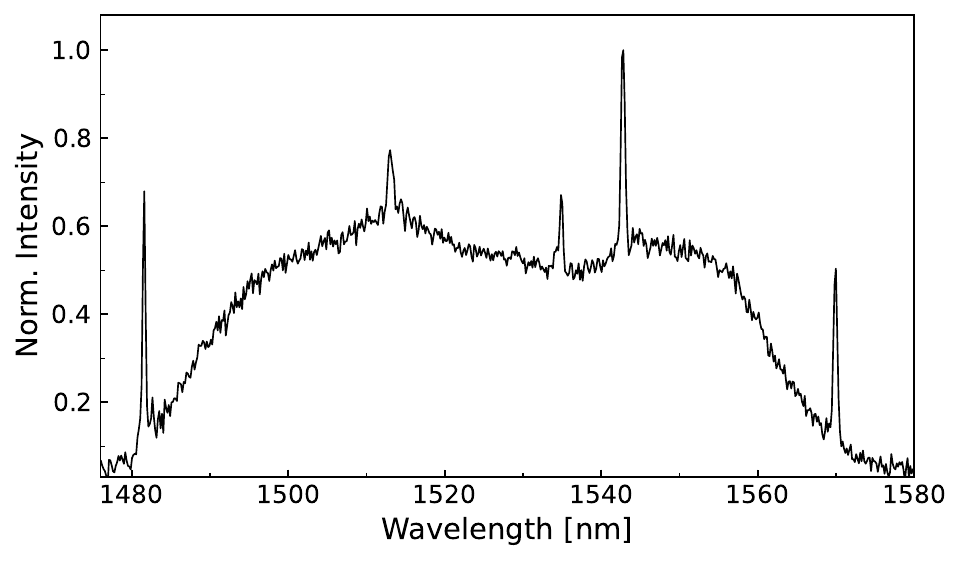}
    \caption{Photoluminescence spectrum of a nanobeam containing ensemble of Al1-centers. The broad band ranging from 1480 nm to 1540 nm is associated to the lattice disorder generated during the implantation process and not fully healed during the thermal annealing step \cite{davies2006radiation}. The sharp peaks are related to Al1-center and its phonon sideband at $\sim$1480 nm and $\sim$1510 nm, and C-center at $\sim$1570 nm.}
    \label{fig:PL_lattice_disorder}
\end{figure}

\textbf{Figure \ref{fig:PL_lattice_disorder}} shows a broad range photoluminescence spectrum of the sample investigated by means of photoluminescence excitation spectroscopy. The spectrum shows some narrow lines associated to known point defects in silicon (Al1-center and its phonon sideband at $\sim$1480 nm and $\sim$1510 nm respectively, and C-center at $\sim$1570 nm) lying on a broad background. This background shape is commonly reported in ensemble measurements of silicon point defects \cite{davies2006radiation} and it is associated to the lattice damage that is unavoidably generated during the defect fabrication process, i.e. ion irradiation or implantation. To create Al1-centers, we employed a single thermal annealing step after both carbon and aluminum implantation at the conditions reported by \cite{irion1988aluminum, irion1989photoluminescence} to give the most intense emission from ensembles. However, identifying the best recipe for single Al1-centers generation will require a further dedicated study of annealing conditions to optimize the Al1-center generation and background reduction. Indeed, aluminum implantation produces $\sim$5 times the amount of vacancies generated by carbon implantation from SRIM (Stopping and Range of Ions in Matter) simulations, due to aluminum bigger mass. Thus it requires different post-implantation annealing conditions to fully recrystallize the induced lattice damage compared to well-studied silicon point defects such as G- and T-centers. At the same time, annealing temperature should favor carbon mobility without going beyond the center thermal stability threshold, which is currently not well-known. 

\subsection*{Power saturation and second-order autocorrelation analysis}

For fitting the two saturation curves shown in \textbf{Figure \ref{fig2}c} we employ a two-level system model, whose analytical expression is given by:
\begin{equation} \label{eq:saturation}
    I(P) = \frac{I_{\mathrm{sat}} \cdot P}{P + P_{\mathrm{sat}}}
\end{equation}
where $I_{\mathrm{sat}}$ and $P_{\mathrm{sat}}$ are saturation count rate and saturation power, respectively.

Background and $g^{(2)}(0)$ correction of the saturation curve has been performed according to the equation:
\begin{equation}\label{saturation_curve_correction}
    I_{\mathrm{corr}} = (I_{\mathrm{meas}}-B) \cdot \sqrt{1-g^{(2)}(0)}
\end{equation}
where $I_{\mathrm{meas}}$ is the measured count rate, $B$ is the background count rate for that incident power, and $g^{(2)}(0)$ is the zero-delay value of the autocorrelation function.

Fitting of the second-order autocorrelation measurements has been performed employing the following equation:

\begin{equation} \label{eq:g2}
    g^{(2)}(\tau) = A \cdot \left( 1 - (1-g^{(2)}(0)) \cdot \exp\left( - \frac{|\tau-t_0|}{\tau_C}\right) \right)
\end{equation}

where A is a pre-factor accounting for imperfect normalization of the experimental data, $g^{(2)}(0)$ is the zero-delay value of the autocorrelation function, $t_0$ is the time offset corresponding to $g^{(2)}(0)$, and $\tau_C$ is the emitter correlation time extracted from exponential decay of the $g^{(2)}(0)$ dip.

\begin{figure}
    \centering
    \includegraphics[width=\columnwidth]{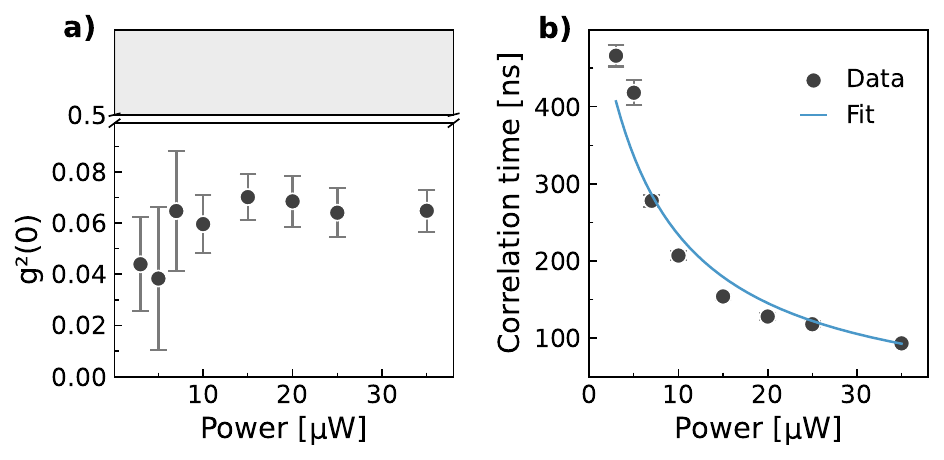}
    \caption{\textbf{a)} Experimental $g^{(2)}(0)$ value as a function of CW 800 nm pump power; \textbf{b)} Correlation time extracted from second-order autocorrelation measurements through fitting by \textbf{Equation \ref{eq:g2}} as a function of CW 800 nm pump power (dots) and corresponding fit (solid line).}
    \label{fig:g2(P)}
\end{figure}

Figure \ref{fig:g2(P)} shows the behavior of $g^{(2)}(0)$ and extracted correlation time $\tau_C$ as a function of above-band laser power. We do not observe a significant change in the single photon purity, while the correlation time gets faster with increasing power as expected. The behavior of $\tau_C$ as a function of pump power can be fitted using the following equation:
\begin{equation}
    \frac{1}{\frac{1}{B} + A \cdot P}
\end{equation}
where A and B are constants and P is the pump power. 

\begin{figure}
    \centering
    \includegraphics[width=\columnwidth]{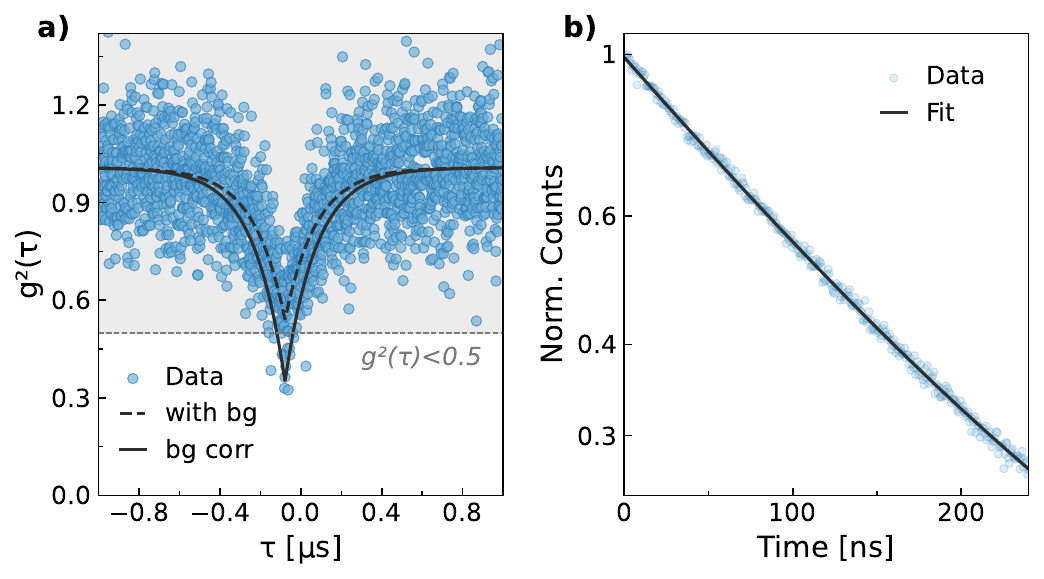}
    \caption{Characterization of the single emitter on which photoluminescence excitation spectroscopy measurements were performed. \textbf{a)} Second-order autocorrelation measurement (dots) and corresponding fit without (dashed line) and with (solid line) background correction. From the latter we extract $g^{(2)}(0) = 0.35 \pm 0.02$; b) Time-resolved photoluminescence measurement (dots) and corresponding single exponential fit (solid line), computing a lifetime of $152 \pm 2$ ns.}
    \label{fig:p2nb4_single}
\end{figure}

Figure \ref{fig:p2nb4_single} shows the photon-correlation and lifetime measurements performed on the same isolated Al1-center used for resonant spectroscopy measurements. The measured photon purity is lower ($g^{(2)}(0) = 0.35 \pm 0.02$) than that of the emitter discussed in the main text because of residual background emission that could not be fully suppressed by spectral filtering. This background may originate both from emission of other Al1-centers less coupled to the nanobeam and from residual implantation induced lattice disorder (see Section \ref{sec:PLE}). Despite this strong background, this center was selected for resonant measurements because its surrounding spectral region is cleaner, enabling clearer resolution of the side transitions, compared to any other single center. The time-resolved photoluminescence decay is well described by a single-exponential model, consistent with emission from an isolated center. This single Al1-center is characterized by a lifetime of $152 \pm 2$ ns.

\subsection*{Magnetospectroscopy via above-band excitation}
\begin{figure*}
    \centering
    \includegraphics[width=\linewidth]{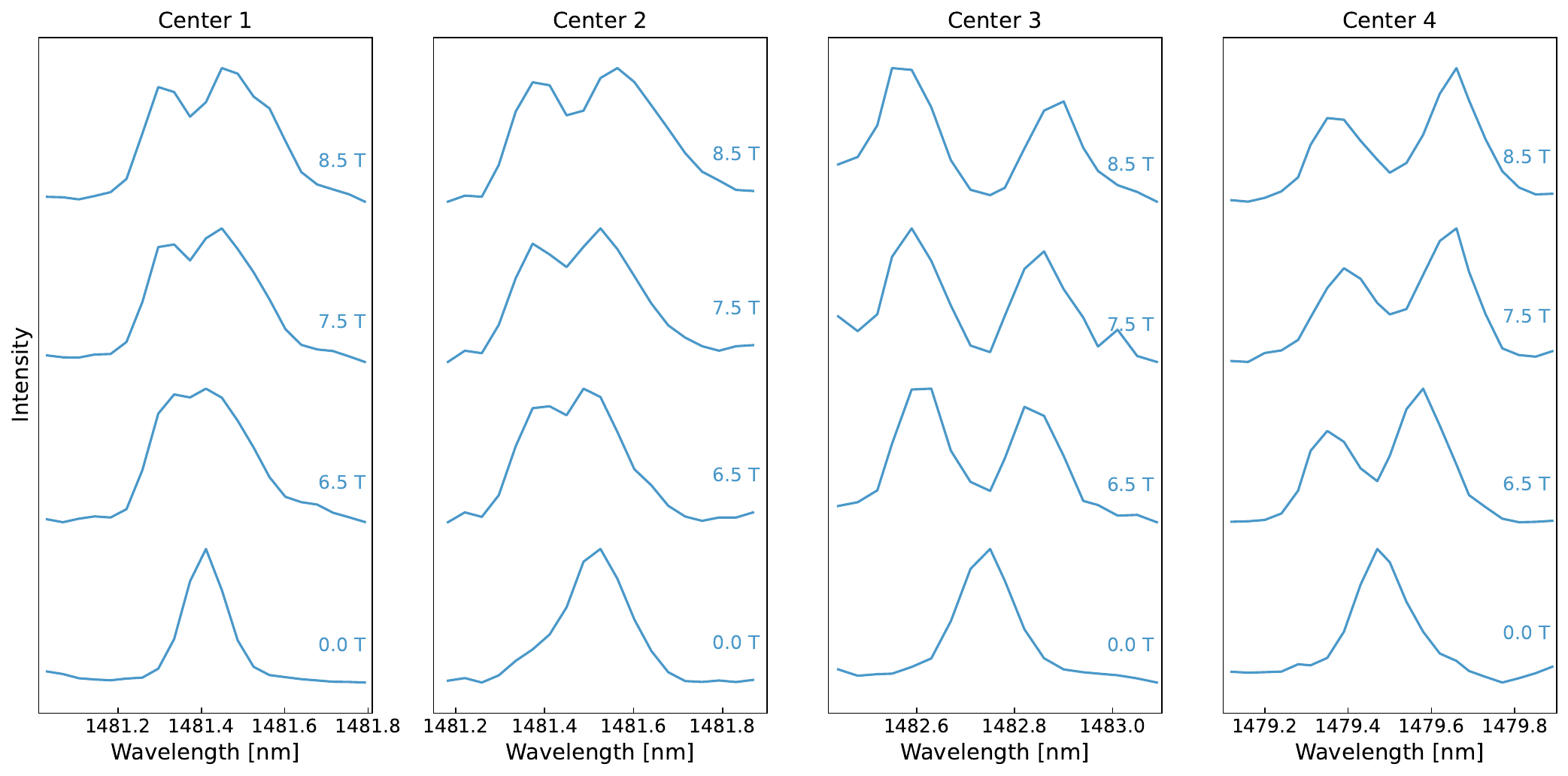}
    \caption{Photoluminescence spectra as a function of external magnetic field for four different single Al1-centers. The extracted $g_{eff}=|g_h-g_e|$ for center 1 to 4 are 0.186 $\pm$ 0.007, 0.216 $\pm$ 0.013, 0.344 $\pm$ 0.006, 0.325 $\pm$ 0.005 respectively.}
    \label{fig:4centers_PL(B)}
\end{figure*}

Before moving to the two-tone resonant measurements, we studied the Zeeman splitting of the zero-phonon line via photoluminescence spectroscopy from a set of single Al1-centers. \textbf{Figure \ref{fig:4centers_PL(B)}} shows the resulting spectra for different magnetic field values. We observe a distribution of effective g-factors extracted from a linear Zeeman fit model, which is in agreement with the known range of hole g-factor values in T-center. In particular, the extracted $g_{eff}=|g_h-g_e|$ values for center 1 to 4 are 0.186 $\pm$ 0.007, 0.216 $\pm$ 0.013, 0.344 $\pm$ 0.006, 0.325 $\pm$ 0.005 respectively.

\section{Resonant excitation measurement}
\label{app}
\setcounter{figure}{0}
\renewcommand{\thefigure}{B\arabic{figure}}
\setcounter{equation}{0}
\renewcommand{\theequation}{B\arabic{equation}}

\subsection*{Extraction of inhomogeneous and homogeneous linewidth}
\begin{figure}
    \centering
    \includegraphics[width=\columnwidth]{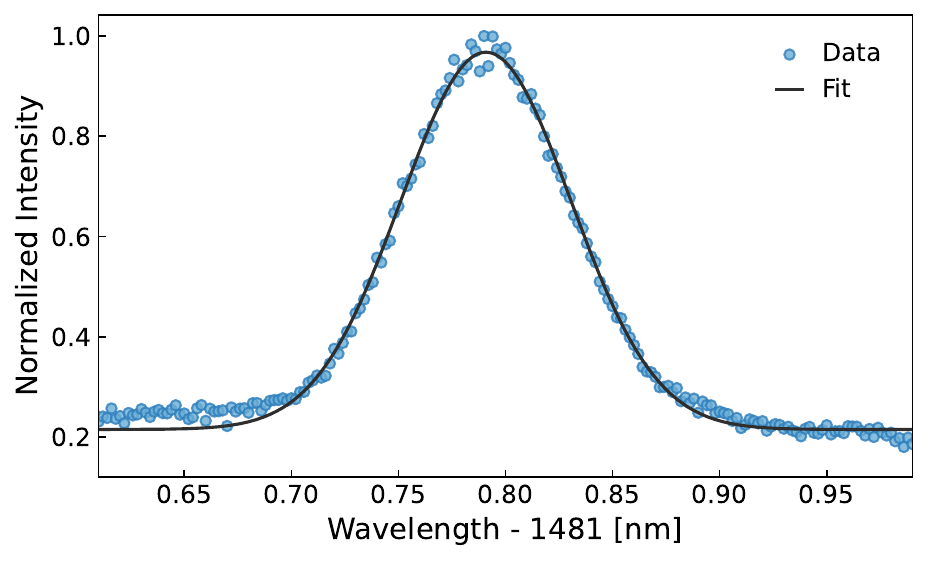}
    \caption{Single color photoluminescence excitation spectroscopy at zero magnetic field (dots) and Gaussian fit of the peak (solid line), extracting an inhomogeneous linewidth of $10.85 \pm 0.08$ GHz.}
    \label{fig:Single_PLE}
\end{figure}

\textbf{Figure \ref{fig:Single_PLE}} shows a single Al1-center zero-phonon line measured via single color photoluminescence excitation spectroscopy at zero magnetic field. The solid line represents the Gaussian fit of the experimental data, from which we extract an inhomogeneous linewidth of $10.85 \pm 0.08$ GHz. 

In two-tone photoluminescence excitation spectroscopy configuration, we perform transient spectral hole burning measurements to extract the homogeneous linewidth of the Al1 defect. The spectral hole full-width at half-maximum is extracted via Lorentz fit of the experimental data. The linewidth behavior with sample temperature reported in \textbf{Figure \ref{fig3}c} is fitted with the following equation \cite{purchase2009spectral, neu2013low}:  
\begin{equation} \label{eq:holeburning_T}
    \Delta \omega_{h}(T) = W_0 + A \cdot T^n
\end{equation}
where $W_0$ is the zero temperature limit of the full-width at half-maximum, $A$ is a constant, and n is a parameter that allows identification of the main homogeneous line broadening mechanism. 
For what concerns power-dependent behavior of the hole width, we use the analytical form \cite{siegman1986lasers}:
\begin{equation} \label{eq:holeburning_P}
    \Delta \omega_{h}(P) = \Delta \omega_{\mathrm{hom}} \left( 1 + \sqrt{1+\frac{P_{\mathrm{probe}}+P_{\mathrm{pump}}}{P_{\mathrm{sat}}^{\mathrm{HB}}}} \right)
\end{equation}
where $P_{\mathrm{probe}}$ and $P_{\mathrm{pump}}$ are the scanning laser and pump laser powers, respectively, and $P_{\mathrm{sat}}^{\mathrm{HB}}$ is the saturation power of the spectral hole. At zero power limit, the hole linewidth reaches twice the homogeneous linewidth. From the fit to experimental data, we extract the spectral hole saturation power to be $P_{\mathrm{sat}}^{\mathrm{HB}} = 0.12 \pm 0.06\ \mathrm{\mu W}$.

\subsection*{Magnetospectroscopy via two-tone resonant excitation}

\begin{figure}
    \centering
    \includegraphics[width=\columnwidth]{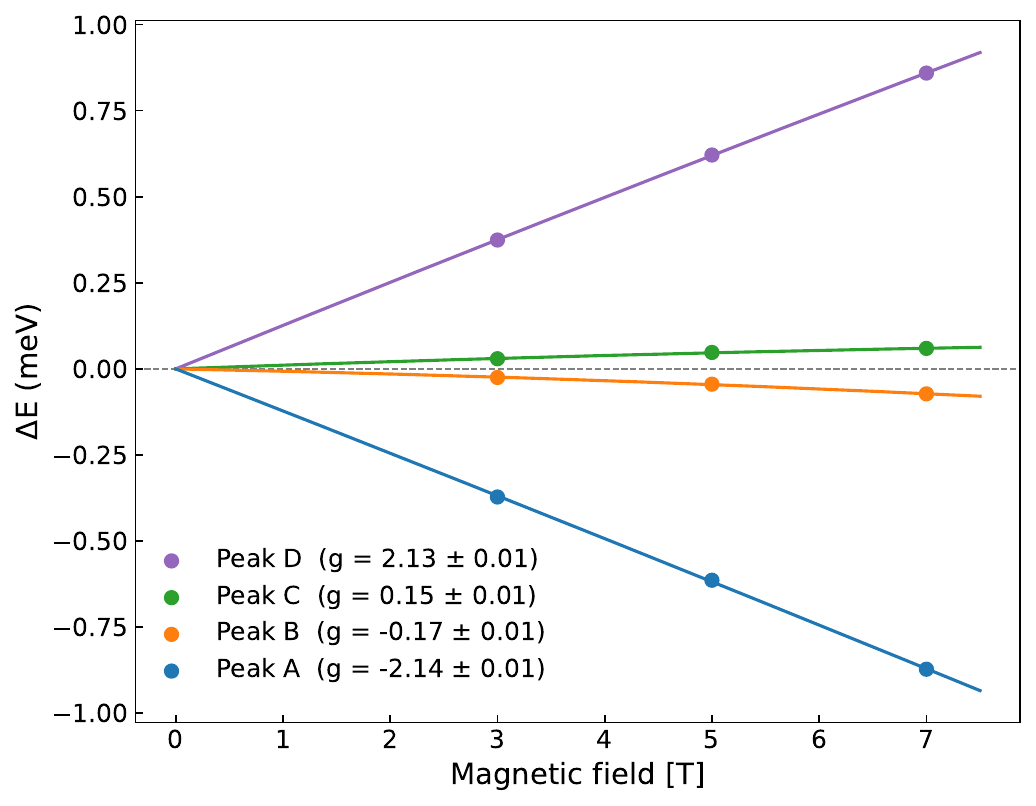}
    \caption{Extracted g-factors from experimental data of \textbf{Figure \ref{fig4}a}.}
    \label{fig:g-factor}
\end{figure}

We extract the peak center energy from \textbf{Figure \ref{fig4}a} and plot the trend of each peak as a function of the applied magnetic field. We fit the resulting behavior to a linear model for Zeeman splitting, extracting the electron and hole g-factors. This is shown in \textbf{Figure \ref{fig:g-factor}}.

\subsection*{Optical pumping and spin-polarization relaxation}

For each resonant power applied to transition C, the fluorescence transient is fitted using:

\begin{equation}
\label{eq:pumping_transient}
I(t)
=
I_{\infty}
+
\left(I_{0}-I_{\infty}\right)
\exp\left(-\gamma_{\mathrm{p}} t\right),
\end{equation}

where $I_{0}$ and $I_{\infty}$ are the initial and steady-state fluorescence levels, respectively, and $\gamma_{\mathrm{p}}$ is the optical-pumping rate. The transient decay results from repeated resonant excitation followed by spontaneous population transfer into the opposite ground-state spin level, which is dark to transition C.

The power dependence of the extracted rate is described by \cite{rosenthal2024single}:

\begin{equation}
\label{eq:pumping_saturation}
\gamma_{\mathrm{p}}(P)
=
\gamma_{\mathrm{p}}^{\mathrm{sat}}
\frac{P}{P+P_{\mathrm{sat}}^{\mathrm{OP}}},
\end{equation}

where $\gamma_{\mathrm{p}}^{\mathrm{sat}}$ is the saturation optical-pumping rate and $P_{\mathrm{sat}}^{\mathrm{OP}}$ is the optical pumping saturation power. Fitting the data in \textbf{Figure~\ref{fig4}c} yields
$\gamma_{\mathrm{p}}^{\mathrm{sat}} = 0.24 \pm 0.05~\mathrm{\mu s}^{-1}$
and
$P_{\mathrm{sat}}^{\mathrm{OP}} = 29 \pm 12~\mathrm{\mu W}$.

In the saturation limit, the excited-state population approaches one half and the pumping rate is related to the optical cyclicity $\eta$ by:

\begin{equation}
\label{eq:cyclicity}
\gamma_{\mathrm{p}}^{\mathrm{sat}}
=
\frac{\Gamma}{2\eta},
\end{equation}

where $\Gamma=1/\tau$ is the excited-state decay rate. The cyclicity is therefore:

\begin{equation}
\label{eq:cyclicity_extraction}
\eta
=
\frac{1}{2\tau\gamma_{\mathrm{p}}^{\mathrm{sat}}}.
\end{equation}

Using $\tau=152\pm2~\mathrm{ns}$, we obtain
$\eta=13.7\pm2.9$, corresponding to an effective spin-changing branching probability of approximately $1/\eta \simeq 7\%$ per optical cycle.

The decay of the optically prepared ground-state spin polarization is determined from the fluorescence contrast measured after a variable dark interval $\Delta t$. For each delay, the contrast is obtained by comparing the integrated fluorescence during the first C readout pulse with that of the final C reference pulse, after subtraction of the background measured during the preceding C pulse. This normalization suppresses slow variations in the excitation and collection efficiencies.

The relaxation of the measured contrast is described by a single-exponential model:

\begin{equation}
\label{eq:contrast}
C(\Delta t)
=
C_{\infty}
+
\left(C_{0}-C_{\infty}\right)
\exp\left(-\frac{\Delta t}{T_{1}}\right),
\end{equation}

where $C_{0}$ is the contrast immediately after optical spin polarization, $C_{\infty}$ is the long-delay contrast reached after relaxation toward thermal equilibrium, and $T_{1}$ is the longitudinal ground-state spin-relaxation time.

\newpage
\bibliographystyle{ieeetr}   % or apalike, plain, unsrt, etc.
\bibliography{refs.bib}          % assumes refs.bib is in the same folder
%\nocite{*}

\clearpage
%\input{sections/Supplementary_Material.tex}

%\appendix*
%\input{sections/appendix1.tex}

\end{document}